\documentclass[11pt]{article}
\usepackage[textwidth=15.2cm,textheight=22cm]{geometry}
\usepackage{amsmath,amssymb}
\usepackage{latexsym}
\usepackage{graphicx}
\usepackage{bm}
\allowdisplaybreaks[1]

\newcommand{\be}{\begin{equation}}
\newcommand{\ee}{\end{equation}}
\newcommand{\ba}{\begin{eqnarray}}
\newcommand{\ea}{\end{eqnarray}}
\newcommand{\bdm}{\begin{displaymath}}
\newcommand{\edm}{\end{displaymath}}

\newcommand\fr[1]{\frac{1}{#1}}

\def\bp{\bar \partial}
\def\ba{\bar A}

\def\beq{\begin{equation}}
\def\eeq{\end{equation}}

\newcommand{\nn}{\nonumber}

\newcommand{\ndt}{\noindent}

\def\bea{\begin{eqnarray}}
\def\eea{\end{eqnarray}}
\def\beas{\begin{eqnarray*}}
\def\eeas{\end{eqnarray*}}
\def\sla{\raise.15ex\hbox{$/$}\kern-.57em}

\def\parm{{\partial}_{-}}

\def\lt{\tilde{\lambda}}
\def\spa#1.#2{\left\langle#1\,#2\right\rangle}
\def\spb#1.#2{\left[#1\,#2\right]}

\begin{document}

\begin{titlepage}
\begin{flushright}    
{\small $\,$}
\end{flushright}
\vskip 1cm
\centerline{\Large{\bf {Spinor helicity structures in higher spin theories}}}
\vskip 1.5cm
\centerline{Sudarshan Ananth}
\vskip .5cm
\centerline{\it {Indian Institute of Science Education and Research}}
\centerline{\it {Pune 411021, India}}
\vskip 1.5cm
\centerline{\bf {Abstract}}
\vskip .5cm
It is shown that the coefficient of the cubic interaction vertex, in higher spin Lagrangians, has a very simple form when written in terms of spinor helicity products. The result for a higher-spin field, of spin $\lambda$, is equal to the corresponding Yang-Mills coefficient raised to the power $\lambda$. Among other things, this suggests perturbative ties, similar to the KLT relations, between higher spin theories and pure Yang-Mills. This result is obtained in four-dimensional flat spacetime.

\vfill
\end{titlepage}

\ndt {\it {Introduction}}
\vskip 0.5cm

\ndt In four-dimensional flat spacetime, Lagrangians describing fields of arbitrary spin have been investigated in detail~\cite{BBB, RRM}. In particular, there exist consistent Lagrangians, to cubic order, for higher spin fields ($\lambda>2$). In this paper, we observe that the coefficient of the cubic interaction vertex in higher spin Lagrangians is equal to the corresponding coefficient in pure Yang-Mills theory raised to the power $\lambda$, ie.
\bea
\label{result}
{\it L^{\,\lambda}_{\,3}}={{\biggl [}\frac{\spa{k}.{l}^{3}}{\spa{l}.{p}\spa{p}.{k}}{\biggr ]}}^\lambda\  ,
\eea
in terms of notation described below. This is shown explicitly in the spin $3$ and spin $4$ cases (and is known already for spins $1$ and $2$) but the structures encountered strongly suggest that the relationship will hold for all spins. This off-shell result, valid at the level of the Lagrangian, is interesting because it suggests that the spinor helicity approach~\cite{PT} that has been so successful in studies of spin $1$ and spin $2$ theories~\cite{SHM} may prove useful in understanding higher spin theories~\cite{HS} as well. For interesting work on related issues see~\cite{MT}.
\vskip 0.3cm
\ndt The result also suggests that there exist KLT-like relations~\cite{KLT} that extend beyond the Yang-Mills -- Gravity system~\cite{MHV, AT}.

\vskip 0.5cm

\ndt {\it {Notation}}

\vskip 0.5cm

\ndt When studying theories based on helicity considerations, it is natural to work in light-cone gauge where only helicity states propagate.  With the metric $(-,+,+,+)$, we define
\be
x^\pm\,=\,\fr{\sqrt 2}\,(x^0\,\pm\,x^3)\ , \quad \partial_\pm\,=\,\fr{\sqrt 2}\,(\partial_0\,\pm\,\partial_3)\ .
\ee
$x^+$ plays the role of light-cone time and $\partial_+$ of the light-cone Hamiltonian. $\parm$ is now a spatial derivative and its inverse, $\frac{1}{\parm}$, is defined using the prescription in~\cite{SM}. We also define
\bea
x&&\!\!\!\!\!\!\!\!=\fr{\sqrt 2}\,(x^1\,+i\,x^2)\ , \quad {\bar \partial}\equiv\frac{\partial}{\partial x}=\fr{\sqrt 2}\,(\partial_1\,-\,i\,\partial_2)\ , \nonumber \\
{\bar x}&&\!\!\!\!\!\!\!\!=\fr{\sqrt 2}\,(x^1\,-i\,x^2)\ , \quad \partial\equiv\frac{\partial}{\partial {\bar x}}=\fr{\sqrt 2}\,(\partial_1\,+\,i\,\partial_2)\ .
\eea
\vskip -0.1cm
\ndt A four-vector $p_\mu$ may be expressed as a bispinor $p_{a \dot{a}}$ using the $\sigma^\mu=(-{\bf {1}},\,{\bf {\sigma}})$ matrices 
\beq
\label{paad}
p_{a \dot{a}}\,\equiv\,p_\mu\,{(\sigma^\mu)}_{a \dot{a}}\,=\,\left( 
 \begin{matrix}
-p_0+p_3 & \;p_1-ip_2\, \\ p_1+ip_2 & \;-p_0-p_3\, 
\end{matrix}
\right)=\sqrt{2}
\left( 
 \begin{matrix}
-p_- & \;{\overline p}\, \\ p & \;-p_+\, 
\end{matrix}
\right)
\ . 
\eeq
The determinant of this matrix is
\bea
{\mbox {det}}\,(\,p_{a \dot{a}}\,)\,=\,-2\,(\,p{\overline p}-p_+p_-\,)\,=\,-\,p^\mu p_\mu\ .
\eea
For light-like $p_\mu$, $p_+\,=\,\frac{p{\overline p}}{p_-}$ represents the on-shell condition, $p^2=0$. We then define holomorphic and anti-holomorphic spinors
\beq
\lambda_{a}\,=\,\frac{2^\fr{4}}{\sqrt p_-} 
\left( 
\begin{matrix} 
p_-  \\ 
\, -p 
\end{matrix}
\right) 
\ , 
\qquad 
\lt_{\dot{a}}\,=\,-(\lambda_a)^*\,=\,-\,\frac{2^\fr{4}}{\sqrt p_-} 
\left( 
\begin{matrix} 
p_- \\ 
\, -{\overline p} 
\end{matrix}
\right) 
\ , 
\eeq
such that $p_{a \dot{a}}=\lambda_{a}\lt_{\dot{a}}$ on-shell. The off-shell holomorphic and anti-holomorphic spinor products read
\be 
\label{product}
\spa{i}.{j}\,=\,\sqrt{2}\,\frac{p^i\,p_-^j\,-\,p^j\,p_-^i}{\sqrt{p_-^i\,p_-^j}}\ ,\qquad \spb{i}.{j}\,=\,\sqrt{2}\,\frac{{\bar p}^i\,p_-^j\,-\,{\bar p}^j\,p_-^i}{\sqrt{p_-^i\,p_-^j}}\  .
\ee

\vskip 0.7cm
\ndt {\it {Higher Spin Lagrangians}}
\vskip 0.5cm

\ndt The light-cone action for a field $\phi$ (and its conjugate $\bar\phi$) of spin $\lambda$, to cubic order, reads~\cite{BBB}
\bea
\label{odd}
S\!=\!\int d^4x\,{\biggl \{}\fr{2}\bar\phi^a\Box\phi^a+\alpha f^{abc} \sum_{n=0}^\lambda {(-1)}^n {\lambda \choose n} \biggl [\bar\phi^a\,\parm^\lambda{\biggl (}\frac{\bp^{(\lambda-n)}}{\parm^{(\lambda-n)}}\phi^b\frac{\bp^n}{\parm^n}\phi^c{\biggr )} +c.c{\biggr ]}\!+\!{\it O}(\alpha^2){\biggr \}}\ ,
\eea
for $\lambda$ odd and
\bea
\label{even}
S\!=\!\int d^4x\,{\biggl \{}\fr{2}\bar\phi\Box\phi+\alpha \sum_{n=0}^\lambda {(-1)}^n  {\lambda \choose n} \biggl [\bar\phi\,\parm^\lambda{\biggl (}\frac{\bp^{(\lambda-n)}}{\parm^{(\lambda-n)}}\phi\frac{\bp^n}{\parm^n}\phi{\biggr )} +c.c{\biggr ]}\!+\!{\it O}(\alpha^2){\biggr \}}\ ,
\eea
for $\lambda$ even. Note that interactions involving fields of odd $\lambda$ are accompanied by antisymmetric structure constants. The Actions above involve three fields all having the same spin and each term involves exactly three derivatives. They thus represent only a subset of all possible cubic interaction Actions involving higher spin fields~\cite{threed}~\footnote{I am grateful to the anonymous referee for pointing this out.}. Since $\phi$ and $\bar\phi$ have definite helicities, both Lagrangians have the following helicity structure
\bea
\label{schemym}
{\it L}\,\sim\,{\it L}_{+-}+\alpha\,{\it L}_{-++}+\alpha\,{\it L}_{+--}+{\it O}(\alpha^2)\  .
\eea
The first cubic vertex in (\ref {schemym}) is non-MHV in structure and could, in principle, be eliminated by a suitable field redefinition if our aim were to produce an MHV Lagrangian~\cite{MHV}. However, since the focus of this paper is on the structure of cubic interaction vertices, this is not necessary. To obtain (\ref {result}), it is sufficient to focus on the ${\it L}_{+--}$ vertex in momentum space.  The $\lambda=1$  and $\lambda=2$ versions of (\ref {result}) are well known so we concentrate on spin $3$ and spin $4$ fields.

\vskip 0.5cm
\ndt {\it {Spin $3$}}
\vskip 0.5cm

\ndt From (\ref {odd}), the cubic interaction vertex for a spin $3$ field in momentum space reads
\bea
\label{spin3}
{\it L}_{+--}^{\lambda=3}=&&\!\!\!\!\!\!\int d^4p\,d^4k\,d^4l\, {(k_-\!+\!l_-)}^3{\biggl \{}\frac{k^3}{2{k_-}^3}-\frac{l^3}{2{l_-}^3}+\frac{3kl^2}{2k_-{l_-}^2}-\frac{3k^2l}{2{k_-}^2l_-}{\biggr \}} \nn \\
&&\times\;\delta^4(p\!+\!k\!+\!l)\,f^{abc}\,\phi^a(p)\bar\phi^b(k)\bar\phi^c(l)\  .
\eea
This formula has been made manifestly antisymmetric in the momenta $k$ and $l$. Recasting (\ref {spin3}) in terms of the off-shell holomorphic spinor product in (\ref {product}) is straightforward and yields
\bea
{\it L}_{+--}^{\lambda=3}=\int d^4p\,d^4k\,d^4l\, {\sqrt 2}\;\frac{\spa{k}.{l}^{9}}{{\spa{l}.{p}}^3{\spa{p}.{k}}^3} \;\delta^4(p\!+\!k\!+\!l)\,f^{abc}\,\phi^a(p)\bar\phi^b(k)\bar\phi^c(l)\  .
\eea

\vskip 1cm
\ndt {\it {Spin $4$}}
\vskip 0.5cm

\ndt The cubic interaction vertex for a spin $4$ field, in momentum space, from (\ref {even}) reads
\bea
\!\!\!\!\!\!\!\!{\it L}_{+--}^{\lambda=4}=&&\!\!\!\!\!\!\!\!\!\!\int d^4p\,d^4k\,d^4l\, {(k_-\!+\!l_-)}^4{\biggl \{}\frac{k^4}{2{k_-}^4}+\frac{l^4}{2{l_-}^4}-\frac{4kl^3}{k_-{l_-}^3}-\frac{4k^3l}{{k_-}^3l_-}+\frac{6k^2l^2}{{k_-}^2{l_-}^2}{\biggr \}} \nn \\
&&\times\;\delta^4(p\!+\!k\!+\!l)\,\phi(p)\bar\phi(k)\bar\phi(l)\  ,
\eea
this expression having been made symmetric in $k$ and $l$. This simplifies to
\bea
\!\!\!\!\!\!\!\!{\it L}_{+--}^{\lambda=4}=\int d^4p\,d^4k\,d^4l\, \fr{4}\;\frac{\spa{k}.{l}^{12}}{{\spa{l}.{p}}^4{\spa{p}.{k}}^4} \;\delta^4(p\!+\!k\!+\!l)\,\phi(p)\bar\phi(k)\bar\phi(l)\  .
\eea
\vskip 0.3cm
\ndt The derivative structures in (\ref {odd}) and (\ref {even}) strongly suggest that the relationship (\ref {result}) holds for all spins, as far as the cubic vertex is concerned.

\vskip 1 cm
\begin{center}
* ~ * ~ *
\end{center}

\ndt An interesting question is whether relations of the type in (\ref {result}) can be established for higher order vertices in the Lagrangian. If so, do they define a consistent interacting tree-level S-matrix with the usual properties?~\footnote{I thank Lance Dixon for this point} For related discussions see~\cite{lit}. Also, do spinor helicity structures simply put in an appearance at order $\alpha$ or are they an indication that higher spin theories possess interesting mathematical structures, similar to those that arise in Yang-Mills theory such as MHV amplitudes. Finally, just as gravity is in some ways the ``square" of Yang-Mills~\cite{KLTfuture}
\beas
{\mbox {Gravity}} \sim ({\mbox {Yang-Mills}}) \times ({\mbox {Yang-Mills}})\ ,
\eeas
formula (\ref {result}) could be an indication that
\beas
({\mbox {Spin}}\!\!-\!\!\lambda) \sim {({\mbox {Yang-Mills}})}^\lambda\ .
\eeas

\vskip 1cm
\ndt {\it {Acknowledgments}}
\vskip 0.3cm

\ndt I thank Hidehiko Shimada, Stefano Kovacs and Stefan Theisen for helpful advice. I also thank Lars Brink and Y. S. Akshay for discussions on related topics. This work is supported by the Max Planck Institute for Gravitational Physics through the Max Planck Partner Group in Quantum Field Theory and the Department of Science and Technology, Government of India, through a Ramanujan Fellowship.

\end{document}